\documentclass[12pt]{article}
\usepackage[applemac]{inputenc}
\usepackage{amsmath}
\usepackage{amsfonts}
\usepackage{amssymb}
\usepackage{mltex}
\usepackage{graphicx}
\newtheorem{lemma}{Lemma}[section]
\newtheorem{theorem}[lemma]{Theorem}
\newtheorem{proposition}[lemma]{Proposition}
\newtheorem{corollary}[lemma]{Corollary}
\newtheorem{remark}[lemma]{Remark}

\newtheorem{definition}[lemma]{Definition}

\def\sq{\hbox {\rlap{$\sqcap$}$\sqcup$}}
\def\sq{\hbox {\rlap{$\sqcap$}$\sqcup$}}

\def\1{{\rm 1\mskip-4.5mu l} }
\def\lsim{\raise0.3ex\hbox{$<$\kern-0.75em\raise-1.1ex\hbox{$\sim$}}}
\def\gsim{\raise0.3ex\hbox{$>$\kern-0.75em\raise-1.1ex\hbox{$\sim$}}}

\def\beq{\begin{equation}}   \def\edq{\end{equation}}
\def\bea{\begin{eqnarray}}  \def\eea{\end{eqnarray}}

\renewcommand{\theequation}{\thesection.\arabic{equation}}
\newcounter{hran} \renewcommand{\thehran}{\thesection.\arabic{hran}}

\def\bmini{\setcounter{hran}{\value{equation}}
    \refstepcounter{hran}\setcounter{equation}{0}
    \renewcommand{\theequation}{\thehran\alph{equation}}\begin{eqnarray}}

\def\bminiG#1{\setcounter{hran}{\value{equation}}
\refstepcounter{hran}\setcounter{equation}{-1}
\renewcommand{\theequation}{\thehran\alph{equation}}
\refstepcounter{equation}\label{#1}\begin{eqnarray}}

\def\emini{\end{eqnarray}\relax\setcounter{equation}{\value{hran}}\renewcommand{
\theequation}
{\thesection.\arabic{equation}}}

\begin{document}
{\title{{\bf \LARGE CIRCULANT MATRICES, GAUSS SUMS AND MUTUALLY
 UNBIASED BASES,\\
  I. THE PRIME NUMBER CASE
}}}
\author{Monique Combescure}
\maketitle

\begin{abstract}
In this paper, we consider the problem of Mutually Unbiased Bases in prime dimension $d$. It is
known to provide exactly $d+1$ mutually unbiased bases. We revisit this problem using a class of
circulant $d \times d$ matrices. The constructive proof of a set of $d+1$ mutually unbiased bases
follows, together with a set of properties of Gauss sums, and of bi-unimodular sequences.
\end{abstract}

\section{INTRODUCTION}
Mutually Unbiased Bases (MUB's) are a set $\left\{\mathcal B_{0}, ..., \mathcal B_{N}\right\}$
of orthonormal bases of $\mathbb C^d$ such that the scalar product in $\mathbb C^d$ of any
vector in $\mathcal B_{j}$ with any vector in $\mathcal B_{k},\ \forall j\ne k$ is of
modulus $d^{-1/2}.$ Starting from the natural base $\mathcal B_{0}$ consiting of vectors
$v_{1}= (1, 0,... 0),\ v_{2}= (0, 1, 0, ..., 0), ... , v_{d}= (0, 0, ... , 1)$, it is known that this
problem reduces to find $N$ unitary Hadamard matrices $P_{j}$ such that $P_{j}^*P_{k}$
is also a unitary Hadamard matrix $\forall j \ne k$. (A unitary matrix is Hadamard if all its
 entries are
of modulus $d^{-1/2}$). The problem has been solved in prime power dimension $d=p^n, \ p,n\in
\mathbb N, p$ prime, and yields exactly $d+1$ MUB's which is the maximum number of
MUB's (\cite{band} and references herein contained).\\
If $d $ is factorizable in $p_{1}^{m_{1}}p_{2}^{m_{2}}...$ with $p_{i}\ne p_{j}$ prime numbers, it is known that
one has at least $N= {\rm min }\ p_{i}^{m_{i}}$ \cite{co}.\\

\noindent
In this paper we show that in prime dimension $d=p$, the Discrete Fourier Transform $F$ together
with a suitable circulant matrix $C$ allow to construct a set of $d+1$ MUB's. In addition this
construction allows to obtain, as a by-product, a set of properties of Gauss sums of the
following form  :
\beq
\label{gauss}
\left\vert \sum_{k=0}^{d-1}\exp \left(\frac{2i\pi}{d}\left[
\frac{lk(k+1)}{2}+jk\right]\right)\right\vert = \sqrt d,\ \quad \forall j 
\in \mathbb F_{d}, l\in \mathbb F_{d}\ {\rm coprime\ with\ d},\ d \ge 3
\edq
 ($\mathbb F_{d}$ is the field of residues
modulo $d$). 
 A direct proof of this property is given in \cite{sa}. Similar results on generalized Gauss Sums appear in \cite{alki}. The definition of $F$ and of circulant matrices is given below. The natural role 
played by circulant matrices in this context is a new result. The circulant unitary matrices are known to be in
one-to-one correspondence with the bi-unimodular sequences $c= \ (c_{1}, c_{2}, ..., c_{d})$ \cite{sabj},
namely sequences such that $\vert c_{j}\vert = \vert (Fc)_{j}\vert=1$, where $F$ is the
discrete Fourier transform. Not surprisingly Gauss sums appear naturally in this context, since
suitable Gauss sequences are examples of bi-unimodular sequences.\\
At the end of this paper, we  consider the case of non-prime dimension, and show that Gauss
sums properties can be deduced in the odd and in the even dimension cases. In a forthcoming
work we shall consider the case of prime power dimensions and show that the theory of block-circulant
matrices with circulant blocks solve the MUB problem in that case.\\

\noindent 
A $d\times d$ matrix is Hadamard if all its entries have equal moduli \cite{ha}, and
$$H^*H=  d \ \1$$
\begin{definition}
A $d\times d$ matrix $H$ is Hadamard if $\vert H_{j,k}\vert\  {\rm is\  constant}\ \forall
j,k=1,...,d$, and $H^*H= d\ \1$. We call $H$ a ``unitary Hadamard matrix'' if
$$\vert H_{j,k}\vert= d^{-1/2},\ \forall j,k=1,...,d,\ {\rm and}\ \sum_{l=1}^dH_{j,l}^*H_{l,k}= \delta_{j,k}$$
\end{definition}

\begin{definition}
A $d \times d$ matrix $C$ is called {\bf circulant} \cite{da}, and denoted
${\rm circ}(c_{0}, c_{1},..., c_{d-1})$, if all its rows and columns are successive
circular permutations of the first. It is of the form
$$C= \begin{pmatrix}c_{0}&c_{d-1}&.&.&c_{1}\\
c_{1}&c_{0}&.&.&c_{2}\\
.&.&.&.&.\\
c_{d-1}&c_{d-2}&.&.&c_{0}\end{pmatrix}$$
\end{definition}

\begin{proposition}
(i) The set $\mathcal C$ of all $d\times d$ circulant matrices is a commutative algebra.\\
(ii) $\mathcal C$ is a subset of {\bf normal matrices}\\
(ii)
Let $V= {\rm circ}(0,0,..., 1)$. Clearly $V^d= \1$. Then $C$ is circular if and only if it commutes with $V$,
and one has for any sequence $c=(c_{1},..., c_{d})\in \mathbb C^d,\ C= {\rm circ}(c_{1},..., c_{d})$ :
$$C= P_{c}(V)= c_{0}\1+ c_{d-1}V+... + c_{1}V^{d-1}$$
where $P_{c}$ is the polynomial
$$P_{c}(x)= \sum_{k=0}^dc_{k}x^{-k}$$
\end{proposition}

Proof :  See \cite{da}\\
For the product of two circulant matrices $C,\ C'$ (that therefore commute with $V$), one has
$$VCC'= CVC'= CC'V$$
which establishes that $CC'$ is indeed circulant.
\\

Moreover it is well-known that there is a close link between the circulant matrices and the discrete Fourier transform.
Namely the latter diagonalizes all the circulant matrices. The Discrete Fourier Transform is
defined by the following $d \times d$ unitary matrix $F$ with matrix elements :
\beq
\label{DFT}
F_{j,k }= d^{-1/2}\exp\left(\frac{2i\pi jk}{d}\right),\quad j,k =0,1,..., d-1
\edq

\begin{proposition}
(i) The circulant matrix $V= {\rm circ}(0,0,...,1)$ is such that
$$F^*VF= U\equiv {\rm diag}(1,\omega, \omega^2, ..., \omega^{d-1})$$
where 
\beq
\label{omega}
\omega= \exp\left(\frac{2i\pi}{d}\right)
\edq
(ii) Let $C= {\rm circ}(c_{0}, c_{1}, ..., c_{d-1})$ be a circulant $d \times d$ matrix. Then
$$F^*CF= \sqrt d\  {\rm diag}(\hat c_{0}, \hat c_{-1},..., \widehat c_{-(d-1)})$$
where
\beq
\label{conjseq}
\hat c_{l}= \frac{1}{\sqrt d}\sum_{k=0}^{d-1}c_{k}\omega^{kl}
\edq

\end{proposition}

Proof : (i) It is enough to check that $v_{k}$, the $k$-th column vector of $F$ which has components
$$(v_{k})_{j}= \frac{\omega^{jk}}{\sqrt d},\quad j,k=0,1, ..., d-1$$
is eigenvector of $V$ with eigenvalue $\omega^{k}$, which is immediate since
$$(Vv_{k})_{j}= \frac{\omega^{k(j+1)}}{\sqrt d}= \omega^k(v_{k})_{j}$$
(ii) One has (Proposition 1.3(ii))
$$C= \sum_{k=0}^{d-1}c_{k}V^{-k}$$
Thus
$$F^*CF= \sum_{k=0}^{d-1}c_{k}(F^*VF)^{-k}= \sum_{k=0}^{d-1}c_{k}U^{-k}= {\rm diag}
(d_{0}, ..., d_{d-1})$$
with
$$d_{l}= \sum_{k=0}^{d-1}c_{k}\omega^{-lk}= \sqrt d \ \hat c_{-l}$$

\begin{lemma}
For any $k\in \mathbb N$, we denote by $[k]$ the rest of the division of $k$ by $d$.
Given any sequence $c= (c_{0}, ..., c_{d-1})\in \mathbb C^d$ its autocorrelation function
obeys
$$\sum_{k=0}^{d-1}\bar c_{k}c_{[j+k]}= \sum_{l=0}^{d-1}\vert \hat c_{l}\vert^2 \omega^{-jl}$$
where the Fourier transform $\hat c$ of $c$ has been defined in (\ref{conjseq}).
\end{lemma}

Proof : For any $j=0, ..., d-1$, one has
$$\sum_{l=0}^{d-1}\vert \hat c_{l}\vert^2\omega^{-jl}= \frac{1}{d}\sum_{l=0}^{d-1}
\omega^{-jl}\sum_{k=0}^{d-1}\bar c_{k}\omega^{-lk}\sum_{k'=0}^{d-1}c_{k'}\omega^{k'l}$$
$$= \sum_{k,k' =0}^{d-1}\bar c_{k}c_{k'}\frac{1}{d}\sum_{l=0}^{d-1}
\omega^{l(k'-k-j)}= \sum_{k=0}^{d-1}c_{[j+k]}\bar c_{k}$$
since
 $$d^{-1}\sum_{l=0}^{d-1}\omega^{l(k'-j-k)}= \delta_{k', [j+k]}$$

\noindent
It is known (\cite{sabj}, \cite{tu}) that {\bf circulant unitary Hadamard matrices} are in one-to-one correspondance
with {\bf bi-unimodular sequences } $c= (c_{0}, c_{1},...,c_{d-1})$.

\begin{definition}
A sequence $c=(c_{0}, c_{1}, ..., c_{d-1})$ is called bi-unimodular if one has
$\vert c_{j}\vert = \vert \hat c_{j}\vert=1, \ \forall j=1,...,d$, where
$\hat c_{j}$ is defined by (\ref{conjseq}).
\end{definition}

\begin{proposition}
Let $(c_{0}, ..., c_{d-1})$ be a bi-unimodular sequence. Then the circulant matrix 
$C= d^{-1/2}{\rm circ}(c_{0}, c_{1},..., c_{d-1})$ is an unitary Hadamard matrix. 
\end{proposition}

Proof : This is a standard ``if and only if'' statement. One uses Lemma 1.5 :
$$\sum_{k=0}^{d-1}\bar c_{k}c_{[j+k]}=
\sum_{l=0}^{d-1}\vert \hat c_{l}\vert^2 \omega^{-jl}$$
But since $\vert \hat c_{l}\vert =1, 
\ \forall l=0,..., d-1,$ the RHS is
simply $d\delta_{j,0}$, and therefore
$$\sum_{k=0}^{d-1}\bar c_{k}c_{[k+j]}=d\  \delta_{j,0}$$
which proves the unitarity of the circulant Hadamard matrix $C$.\\

\noindent
In all that follows we call indifferently $F$ or $P_{0}$ the discrete Fourier transform.\\

\noindent
In \cite{klimuro}, the authors introduce for any dimension $d$ being the power of a prime number
a set of operators called ``rotation operators'' which can be viewed as ``circulant matrices''
(this property is however not put forward explicitely by the authors). In this paper, restricting
ourselves to the {\bf prime number case}, we show that these operators can be used to define
a set of $d+1$ Mutually Unbiased Bases in dimension $d$.\\

Mutually Unbiased bases are extensively studied in the framework of Quantum Information 
Theory. They are defined as follows :

\begin{definition}
A set $\left\{ B_{1},B_{2},...,B_{m}\right\}$ of orthonormal bases of $\mathbb C^d$
is called MUB if for any vector $b_{j}^{(k)}\in B_{k}$ and any $b_{j'}^{(k')}\in B_{k'}$
 one has
$$\vert b_{j}^{(k)}\cdot b_{j'}^{(k')}\vert= d^{-1/2},\ \forall k\ne k'= 1,...,m,\ 
\forall j,j'=1,...,d$$
where the dot represents the Hermitian scalar product in $\mathbb C^d$.
\end{definition}

\begin{remark}
It is trivial to show that if the orthonormal bases $B_{k}$ are the column vectors of an unitary
matrix $A_{k}$, then the property that must satisfy the $A_{k}$'s in order that 
$\left\{\1, B_{1},...,B_{m}\right\}$ are MUB's is that all 
$A_{k},\ k=1,...,m$ and $A_{k}^*A_{k'}, \ k\ne k'=1,...,m$ are
unitary Hadamard matrices. Namely if $u_{j},\ v_{k}$ are column vectors for  unitary matrices
$A,\ A'$ respectively, then
$$u^{j}\cdot v^{k}=(A^*A')_{j,k} $$
Thus {\bf unitary Hadamard matrices} play a major role in the MUB problem.
\end{remark}

It is known that the maximum number of MUB's in any dimension $d$ is $d+1$, and that this number
is attained if $d=p^m$, $p$ being a prime number. In this paper, restricting ourselves to $m=1$,
we revisit the proof of this property, using circulant matrices introduced by \cite{klimuro}.
We then show that it implies beautiful properties of Gauss sums, namely the following (\cite{sa}) :\\

\begin{proposition}
Let $d\ge 3$ be an odd number. Then $\forall k=1, ..., d-1$ coprime with $d$ the sequences
\beq
\label{gauss1}
g^{(k)}:= \left(\exp\left(\frac{i\pi kj(j+1)}{d}\right)\right)_{j=0, ..., d-1}
\edq
are bi-unimodular. Thus (\ref{gauss}) holds true.

\end{proposition}

This  property will appear as a subproduct of our study of MUB's for $d$ a prime number $\ge 3$
via the circulant matrices method. As stressed above, the link between circulant matrices and 
bi-unimodular sequences is well established. What is new here is the fact that the MUB problem
via a circulant matrix method allows to recover the bi-unimodularity of Gauss sequences. The 
crucial role played by the Gauss sequence is due to the crucial role played by the Discrete
Fourier Transform (or in other therms the Fourier-Vandermonde matrices) in the MUB problem
for prime numbers. Let us introduce it now explicitely.\\

\noindent
It is known since Schwinger (\cite{sch}) that a simple toolbox of unitary $d\times d$ matrices sometimes
refered to as ``generalized Pauli matrices'' $U,V$ allows to find MUB's. $U,\ V$ generate the 
discrete Weyl-Heisenberg group \cite{we}.
 Denote by $\omega$ the primitive root of unity (\ref{omega}).
The matrix $U$ is simply
$$U= {\rm diag}(1,\omega,\omega^2,...,\omega^{d-1})$$
which generalizes the Pauli matrix $\sigma_{z}$ to dimensions higher than two. The matrix
$V$ generalizes $\sigma_{x}$ :
$$V={\rm circ}(0, 0, ..., 1)=\begin{pmatrix}0&1&0&.&.&0\\
0&0&1&.&.&0\\
.&.&.&.&.&.\\
0&0&0&.&.&1\\
1&0&0&.&.&0\end{pmatrix}$$
Then one has the following result :

\begin{theorem}
(i) The $U,\ V$ matrices obey the $\omega$-commutation rule :
$$VU=\omega UV$$
(ii) The Discrete Fourier Transform matrix $P_{0}=F$ defined by (\ref{DFT}), namely
$$P_{0}= \frac{1}{\sqrt d}\begin{pmatrix}1&1&1&.&.&1\\
1&\omega&\omega^2&.&.&\omega^{d-1}\\
1&\omega^2&\omega^4&.&.&\omega^{2(d-1)}\\
.&.&.&.&.&.\\
1&\omega^{d-1}&\omega^{2(d-1)}&.&.&\omega^{(d-1)(d-1)}\end{pmatrix}$$
diagonalizes $V$, namely
$$V=P_{0}UP_{0}^*= P_{0}^*U^*P_{0}$$
(iii) One has $P_{0}^4= \1$
\end{theorem}

(ii) is simply Proposition 1.4(i).
For the proof of (iii) reminiscent to the properties of continuous Fourier transform, it is enough
to check that
$$P_{0}^2= W = \begin{pmatrix}1&0&0&0&.&.&0\\
0&0&0&0&.&.&1\\
.&.&.&.&.&.&.\\
.&.&.&.&.&.&.\\
0&0&1&0&.&.&0\\
0&1&0&0&.&.&0\end{pmatrix}$$
Thus $W=\1$ in dimension $d=2$, and $W^2=\1,\quad \forall d\ge 3$.\\

\noindent
In \cite{co} we have shown that for $d$ odd one can add to the general toolbox of unitary
Schwinger matrices $U,\ V$ a diagonal matrix $D$ of the form
$$D={\rm diag}(1,\omega,\omega^3,...,\omega^{k(k+1)/2}, ...,1)$$ so that the MUB problem
for odd prime dimension reduces to properties of $U,\ V, \ D$, and that certain properties
of quadratic Gauss sums follow as a by-product.

\section{THE d=2 CASE}

We have $U=\sigma_{z}$ and $V= \sigma_{x}$, $\sigma_{z},\ \sigma_{x}$ being the
usual Pauli matrices. Since $UV= \sigma_{y}$, finding MUB's in dimension $d=2$ amounts
to diagonalize $\sigma_{x},\ \sigma_{y}$. One has :
$$\sigma_{x}= P_{0}\sigma_{z}P_{0}^*$$
$$\sigma_{y}=P_{1}\sigma_{z}P_{1}^*$$
with
$$P_{0}= \frac{1}{\sqrt 2}\begin{pmatrix}1&1\\
1&-1\end{pmatrix},\quad P_{1}= \frac{1}{\sqrt 2}\begin{pmatrix}1&i\\
i&1\end{pmatrix}$$
$P_{1}$ is circulant.

\begin{proposition}
The set $\left\{ \1, P_{0},P_{1}\right\}$ defines three MUB's in dimension 2.
\end{proposition}

Since the matrices $P_{0},\ P_{1}$ are trivially unitary Hadamard matrices, it is enough to check that $P_{0}^*P_{1}$
is itself a unitary Hadamard matrix, which holds true since
$$P_{0}^*P_{1}= \frac{e^{i\pi/4}}{\sqrt 2}\begin{pmatrix}1&1\\
-i&-i\end{pmatrix}$$

\section{THE PRIME DIMENSION $d\ge 3$}

$d$ being prime, we denote by $\mathbb F_{d}$ the Galois field of integers mod $d$. \\
Let us recall the definition of the ``rotation operator'' of \cite{klimuro}, which, as already stressed
in nothing but a {\bf circulant matrix} in the odd prime dimension $d$.

\begin{definition}
Define $R$ as an unitary operator commuting with $V$ and diagonalizing $VU$.

\end{definition}

\begin{proposition}
(i) $R$ is a circulant matrix.\\
(ii)  $R^k$ is also circulant $\forall k\in \mathbb Z$.
\end{proposition}
This follows from Proposition 1.3.
\\

\noindent
Therefore we are led to consider a subclass of circulant matrices that are unitary. They must
satisfy :\\
$\forall k=0,...,d-1 \quad \vert c_{k}\vert = d^{-1/2}$ and\\
$\forall k=1,...,d-1\quad \sum_{j=0}^{d-1}\bar c_{j}c_{[d-k+j]}=0$ (orthogonality condition).\\

Now it remains to show that such a matrix $R $ exists. In \cite{co} we have constructed a unitary
matrix $P_{1} $ that diagonalizes $VU$. It is defined as
$$P_{1}= D^{-1}P_{0}$$
for any $d$ odd integer (not necessarily prime).
\\
We have established the following result :

\begin{proposition}
(i) For any odd integer $d$,  the matrix $P_{0}^*P_{1}$ is a unitary Hadamard matrix.\\
(ii) For $d\ge 3$ odd integer, let $P_{k}:= D^{-k}P_{0}$. 
Then $P_{0}^*P_{k}$ is a unitary Hadamard
matrix for all $k$ coprime with $d$.\\
(iii) The matrix $D$ defined above is such that 
$$\vert{\rm Tr}D^k\vert=\sqrt d,\ \forall k\in \mathbb F_{d}\ {\rm coprime\ with\ }d$$
\end{proposition}

For the simple proof of this result see \cite{co}. (iii) is a simple consequence of (i) and(ii). Namely
the  eigenvector of $VU$ belonging to the eigenvalue 1 has components
$$v_{j}=d^{-1/2} \omega^{-\frac{j(j+1)}{2}}$$
Since $P_{0}^*P_{1}$ is unitary Hadamard matrix, the  element of $P_{0}^*P_{1}$ of
the first row and first column is simply
$$d^{-1}\sum_{j=0}^{d-1}\omega^{-\frac{j(j+1)}{2}}$$
and since its modulus must be $d^{-1/2}$ we obtain (iii) for $k=1$. The proof for any $k$ 
coprime with $d$ can be obtained similarly using (ii).
\\

\noindent
The ``problem'' is that $P_{1}$ is not circulant. However the circulant matrix $R$ is obtained from
$P_{1}$ by multiplying the $k$th column vector of $P_{1}$ by a phase which is
$$\omega^{-\frac{k(k-1)}{2}}$$
This operation preserves the fact that it is unitary and that it diagonalizes $VU$. We thus have :
\beq
\label{diag}
VU= P_{1}UP_{1}^*= RUR^*
\edq

\begin{proposition}
Let $R$ be the  matrix :
$$R= d^{-1/2}{\rm circ}\ (1,\omega^{-1},\omega^{-3},...,\omega^{-k(k+1)/2},...,1)$$
It is a unitary Hadamard matrix.
\end{proposition}

Proof : By construction it is a unitary Hadamard matrix, since $P_{1}$ is. To prove the
 fact that it is circulant, it is enough to know that
$$(P_{1})_{j,k}= d^{-1/2}\omega^{jk - \frac{j(j+1)}{2}}$$
thus
\beq
\label{elemmatrix}
R_{jk}= d^{-1/2}\omega^{jk-\frac{j(j+1)}{2}- \frac{k(k-1)}{2}}= 
d^{-1/2}\omega^{-\frac{(j-k)(j-k+1)}{2}}
\edq
since 
$$jk-\frac{j(j+1)}{2}- \frac{k(k-1)}{2}= -\frac{(j-k)(j-k+1)}{2}$$
Thus all column vectors are obtained from the first by the circularity property.
\\

\noindent
Furthermore the $R^k$ have the property that  they diagonalize
 $V^kU,\ \forall k=1,...,d-1$:
 
 \begin{theorem}
 (i) $R= \alpha P_{0}DP_{0}^*$ where
 $\alpha:= d^{-1/2}\sum_{k=0}^{d-1}\omega^{-k(k+1)/2}$ is a complex number
  of modulus 1.\\
  (ii) $R^d= \alpha^{d}\1$ where $\1$ denotes the unity $d \times d $ matrix.\\
 (iii) $$R^kU(R^*)^k= V^kU, \ \forall k=0,...,d$$
 \end{theorem}
 
 \noindent
 The proof is extremely simple :\\
 (i) We have shown that $c_{j}= \omega^{\frac {-j(j+1)}{2}}$ is a bi-unimodular sequence,
  thus $\alpha$ is a complex number of modulus one. 
 Moreover from Proposition 1.4 (ii),  the unitary matrix $R= d^{-1/2}{\rm circ}(c_{j})$ is such that
 $$\hat R =P_{0}^*RP_{0}=  {\rm diag}(\hat c_{-k})$$
 But 
 $$\hat c_{-k}= \frac{1}{\sqrt d}\sum_{j=0}^{d-1}\omega^{-jk-\frac{j(j+1)}{2}}=
 \alpha \omega^{\frac{k(k+1)}{2}}$$
 since
$$\sum_{j=0}^{d-1}\omega^{-\frac{j(j+1)}{2}}= \sum_{j=0}^{d-1}
\omega^{-\frac{(j+k)(j+k+1)}{2}}= \sum_{j=0}^{d-1}\omega^{-jk-\frac{j(j+1)}{2}-
\frac{k(k+1)}{2}}$$
Therefore
$$\hat R=  \alpha  {\rm diag}(\omega^{\frac{k(k+ 1)}{2}})=  \alpha   D$$
\\
(ii) is simply a consequence of (i) since $D^d=\1$.\\
(iii) is obtained by recurrence. Namely it is true for $k=1$ by (\ref{diag}). For $k \ge 2$ one has :
$$R^kU(R^*)^k= RV^{k-1}UR^*= V^{k-1}RUR^*$$
since $R$ commutes with $V$. But
$$V^{k-1}RUR^*= V^{k-1}(VU)= V^kU$$

\noindent
There is a direct link between the matrices $P_{k}$ that diagonalize $VU^k$ and the $R^k$ that
diagonalize $V^kU$ :

\begin{theorem}
$\alpha$ being the complex number of modulus one defined above, we have for any $k=0,...,d-1$
$$P_{k}= \alpha^{k}P_{0}^* R^{-k}P_{0}^2$$
\end{theorem}
 Proof : In \cite{co} we have proven that $P_{k}=D^{-k}P_{0}$ diagonalizes $VU^k$.
 But one has
 $$D^{-k}= D^{d-k}= \alpha^{k-d}P_{0}^*R^{d-k}P_{0}$$
 so the result follows immediately.
 
 \begin{corollary}
  For any $k=1,...,d-1$, $R^k$ is a unitary Hadamard circulant matrix when $d\ge 3 $ is  prime.
 \end{corollary}
 
 Proof : $R^k$ is circulant and unitary since $R$ is. Therefore we have only to check that it is
 Hadamard. We have :
 $$R^{-k}= \alpha^{-k}P_{0}^2P_{0}^*P_{k}P_{0}^2$$
 But we recall that $P_{0}^2$ equals the permutation matrix $W$. Thus all matrix elements of
 $R^k$ equal, up to a phase, some matrix elements of $P_{0}^*P_{k}$. But we have
 established in \cite{co} that the matrix $P_{0}^*P_{k}$ is unitary Hadamard $\forall k=1, ..., d-1$, thus all its
 matrix elements are equal in modulus to $d^{-1/2}$. This completes the proof.
 \\
 
 Now we show how this property of the matrix $R$ reflects itself in Gauss sums properties.
 
 \begin{proposition}
(i)  Let $d$ be an odd prime. Then for any $k=1, 2, ..., d-1$ $R^k$ is an unitary Hadamard matrix
 if and only if one has
$$\left\vert \sum_{j=0}^{d-1}\exp\left(\frac{i\pi}{d}[kj^2+j(k+2m)]\right)\right\vert
 =\sqrt d,\ \forall m=-d+1, ..., d-1$$
 (ii) Under the same conditions $R^k$ is a unitary Hadamard matrix if and only if
 $$\left\vert \sum_{j=0}^{k-1}\exp\left(\frac{i\pi}{k}dj^2+(k+2m)j\right)\right\vert
 =\sqrt k$$
 \end{proposition} 
 
 Proof : Since $R^k= \alpha^k P_{0}D^kP_{0}^*$, the matrix elements of $R^k$ are
 $$(R^k)_{m,l}= \frac{\alpha^k}{d}\sum_{j=0}^{d-1}\omega^{j(m-l)+k\frac{j(j+1)}{2}}$$
 Since $R^k$ is circulant, it is unitary Hadamard matrix if and only if the matrix elements of
 the first column (l=0) are of modulus $d^{-1/2}$, thus if and only if
 $$\left\vert\sum_{j=0}^{d-1}\exp\left(\frac{i\pi}{d}[kj^2+j(k+2m)]\right)\right\vert= \sqrt d$$
 which proves (i).\\
 (ii) Using the reciprocity theorem for Gauss sums (\cite{berndt}), we have for all integers $a,b$
 with $ac\ne 0$ and $ac+d$ even that the quantity
 $$S(a,b,d):= \sum_{j=0}^{d-1}\exp\left(\frac{\pi i}{d}(aj^2+bj)\right)$$
 obeys
 $$S(a,b, d)= \left\vert \frac{d}{a}\right\vert^{1/2}\exp\left( \frac{\pi i}{4}
 [{\rm sgn}(ad)- b^2/ad]\right)S(-d,-b,a)$$
 Thus $\vert S(a,b,d)\vert = \sqrt d$ if and only if
 $\vert S(-d,-b,a)\vert= \sqrt a$. Applying if to $a=k=1,...,d-1$ coprime with $d$ and
 $b=2m+k$, and taking the complex conjugate yields the result. Namely 
 $ad+b= dk+k+2m$ is even for all $k=0,...,d-1$ since $d$ is odd.
 
 \begin{corollary}
 Let $d\ge 3$ be a prime number. Then for any $k=1,2, ..., d-1$ the sequences 
$$g^{(k)}:= \left(\omega^{k\frac{j(j+1)}{2}}\right)_{j=0, ..., d-1}$$
are bi-unimodular.
 \end{corollary}
 
 \begin{remark}
 This property is known, but has an extension in the non prime odd dimensions.
  See next section.
 \end{remark}
 
 It is known that the diagonalization of $VU^k,\ k=0,...,d-1$ provides a set of 
 $d+1$ MUB's for a prime
  $p$. Here we show that the same is true with the diagonalization of $V^kU,\ k=0,...,d-1$.
  
  \begin{theorem}
  Let $d\ge 2$ be a prime dimension. Then the orthonormal bases defined by the unitary
   matrices $\1, P_{0}, R,R^2,...,R^{d-1}$
  provide a set of $d+1$ MUB's.
  \end{theorem}
  
  Proof : For $d=2$ this has been already proven in Section 2. For $d\ge 3$ (thus odd, since it is prime), it
  is enough to check that :\\
  (i) $P_{0}, R^k,\ k=1,...,d-1$ are unitary Hadamard matrices, together
  with\\
  (ii) $P_{0}^*R^k,\ k=1,...,d-1$ and $(R^{k'})^*R^k, \ 1\le k'<k\le d-1$.\\
  Since (i) has been already established, it remains to show (ii). Since
  $R^k= \alpha^kP_{0}D^kP_{0}^*$ we have
  $$P_{0}^*R^k= \alpha^kD^kP_{0}^*$$
  which is trivially an unitary Hadamard matrix (since $P_{0}$ is, $\alpha$ is of modulus 1
  and $D$ is diagonal and unitary). For $(R^*)^{k'}R^k$ it is trivial since
  $$(R^*)^{k'}R^k= R^{k-k'}$$
  which is unitary Hadamard for any $k\ne k',\ k,k'=1,...,d-1$.

  \section{THE CASE OF ARBITRARY ODD DIMENSION}
  
  We have shown in \cite{co} that for any odd dimension $d \ge 3$ the matrices $P_{0}^*P_{k}$
  is an unitary Hadamard matrix {\bf provided $k $ is co-prime with $d$}. This can be transfered
  to a similar property for the matrix $R^{-k}$, and therefore to the bi-unimodularity of
  the sequence $g^{(k)}$ defined in (\ref{gauss1}).
  
  \begin{proposition}
  Let $d\ge 3$ be an odd integer, and $k$ be any number coprime with $d$. Then\\
  (i) $R^k$ is an unitary Hadamard matrix.\\
  (ii) The sequence $g^{(k)}$ is bi-unimodular.\\
  (iii) Both properies are equivalent.
  \end{proposition}
  
  This implies Proposition 1.10.
  
  \begin{theorem}
 Let $d$ be odd and $k>2$ be the smallest divisor of $d$. Then the orthonormal bases defined by the
 unitary matrices $\left\{ \1, F, R, R^2, ..., R^{k-1}\right\}$ provide a set of $k+1$
 MUB's in dimension $d$.
  \end{theorem}
  
  \section{THE CASE OF ARBITRARY EVEN DIMENSION}
  Let $d$ be even and
  $$\omega= \exp\left(\frac{2i\pi}{d}\right)$$
  We denote $\omega^{1/2}= e^{i\pi/d}$.
  One defines the Discrete Fourier Transform $F$ as usually.\\
  The theory of circulant matrices is also pertinent for even dimensions $d\ge 4$.  Namely in that case
  the matrix 
  $$D'= {\rm diag}(1, \omega^{-1/2}, ..., \omega^{-k^2/2}, ..., \omega^{-1/2} )$$
  has been shown (\cite{co}) to be such that the unitary Hadamard matrix
  $$P_{1}= D'F$$
  diagonalizes $VU$. But the circulant matrix $R$ obtained by multiplying the $k$-th column
  vector of $P_{1}$ by $\omega^{-k^2/2}$ also diagonalizes $VU$ :
  
  \begin{proposition}
  The circulant matrix whose matrix elements are
  $$R_{j,k}= \frac{1}{\sqrt d}\omega^{-(j-k)^2/2}$$
  diagonalizes $VU$ and is such that
  $F^*R$ is an unitary Hadamard matrix.
  \end{proposition}
  
  Proof :$$(P_{1})_{j,k}= \frac{1}{\sqrt d}\omega^{-\frac{j^2}{2}+jk}$$
  thus
  $$R_{j,k}= \frac{1}{\sqrt d}\omega^{-\frac{(j-k)^2}{2}}$$
  $R_{j,k}$ only depends on $j-k$ thus is circulant (and unitary). Therefore it is diagonalized by
  $F$, namely there exists an unitary diagonal matrix $D''$ such that
  $$F^*RF=D''$$
  Again this implies that $F^*R$ is an unitary Hadamard matrix.\\
  \sq
  
  \begin{corollary}
  (i) The orthonormal bases defined by the unitary matrices $\left\{ \1, F, R\right\}$
  provide a set of 3 MUB's in arbitrary even dimension $d$.\\
  (ii) One has the following property of quadratic Gauss sums for $d$ even  :
  $$\left\vert\sum_{k=0}^{d-1}\exp\left(\frac{ik^2\pi}{d}\right)\right\vert
  = \sqrt d$$
  \end{corollary}
  
  \begin{remark}
   $R^2$ is circulant,  unitary, but not Hadamard. Thus it does not help to find more than 3 MUB's in even
  dimensions. In dimensions $d=2^n$, another method is necessary to prove that there exists
  $d+1$ MUB's.
  \end{remark}
  
  \medskip
  
  \noindent
  {\bf Acknowledgments :} It is a pleasure to thank B. Helffer for his interest in this work, and to
  Bahman Saffari for interesting discussions and for providing to me reference \cite{sa}.

\end{document}